\documentclass[prl, twocolumn, floatfix]{revtex4}

\usepackage{graphicx}
\usepackage{amsfonts}
\usepackage{wasysym}

\begin{document}

\title{Modeling spin transport in electrostatically-gated lateral-channel silicon devices: role of interfacial spin relaxation}

\author{Jing Li}
\author{Ian Appelbaum}
\altaffiliation{appelbaum@physics.umd.edu}
\affiliation{Center for Nanophysics and Advanced Materials and Department of Physics, University of Maryland, College Park MD 20742 USA}

\begin{abstract}
Using a two-dimensional finite-differences scheme to model spin transport in silicon devices with lateral geometry, we simulate the effects of spin relaxation at interfacial boundaries, i.e. the exposed top surface and at an electrostatically-controlled backgate with SiO$_2$ dielectric. These gate-voltage-dependent simulations are compared to previous experimental results and show that strong spin relaxation due to extrinsic effects yield an Si/SiO$_2$ interfacial spin lifetime of $\approx$ 1ns, orders of magnitude lower than lifetimes in the bulk Si, whereas relaxation at the top surface plays no substantial role. Hall effect measurements on ballistically injected electrons gated in the transport channel yield the carrier mobility directly and suggest that this reduction in spin lifetime is only partially due to enhanced interfacial momentum scattering which induces random spin flips as in the Elliott effect. Therefore, other extrinsic mechanisms such as those caused by paramagnetic defects should also be considered in order to explain the dramatic enhancement in spin relaxation at the gate interface over bulk values.
\end{abstract}

\maketitle

In nonmagnetic materials, a nonequilibrium spin polarization will relax away over a characteristic timescale called the ``spin lifetime''. If the material has weak spin-orbit interaction, dilute nuclear spin species, and lattice spatial inversion symmetry, the conduction electron spin is relatively insulated from other angular momentum degrees of freedom and the relaxation rate is low, resulting in long spin lifetime\cite{YAFET, FABIANWU, DERYLI}. Silicon (Si) has these properties and the predicted long conduction electron spin lifetime in the bulk has been confirmed with magnetic resonance linewidth measurements\cite{LEPINE} and in recent transport experiments probing spin time-of-flight with spin precession\cite{BIQINPRL, LARMORPRBRC}.

\begin{figure}
\includegraphics[width=7.75cm, height=8.75cm]{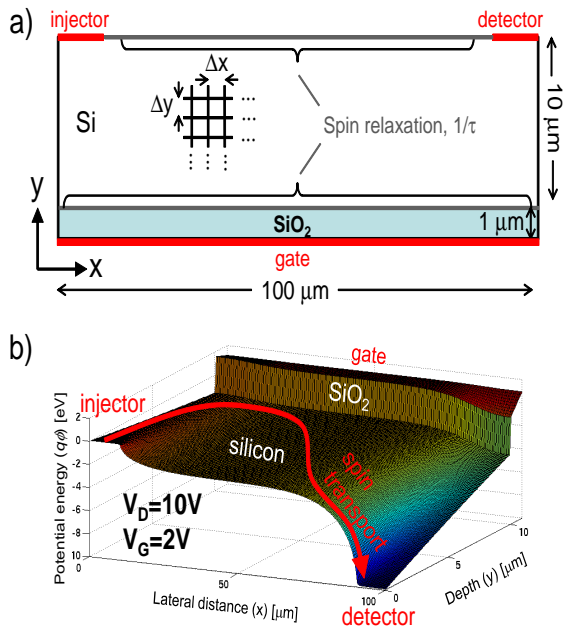}
\caption{\label{FIG1} 
(a) Finite-difference computational domain showing location of spin injector, detector, and gate. (b) Example electrostatic potential landscape (solution of Poisson Eqn. \ref{POISSON} plus 3.1eV conduction band offset between Si and SiO$_2$),\cite{SIO2BARRIER} through which we model spin-polarized transport according to Eqn. \ref{DRIFTDIFFEQ}. Here, $V_G$=2V and $V_D$=10V relative to the injector potential. Spin relaxation is nonzero only at either the top or bottom interfaces as explained in the text; insulating boundary conditions are used except at the detector where spin flows out to consititute a spin current, matching experimental conditions in Ref. \onlinecite{LATERALPRL}.}
\end{figure}

This nearly spin-conserving property of bulk Si is encouraging for the development of applications such as schemes to utilize spin to encode and process information.\cite{DATTANATURENANO,DERYNATURE, SISPINCOMM} However, the spin-enabled devices and circuits necessary in any implementation will invariably require surfaces and interfaces where the circumstances preserving spin in the bulk no longer apply. For instance, paramagnetic defects,\cite{YABLONOVITCH} inversion asymmetry,\cite{DYAKONOV} and fluctuating electric fields,\cite{SHERMAN} all of which are absent in the ideal bulk, are present at interfaces and can drastically increase the spin relaxation rate resulting in a much shorter spin lifetime.\cite{LYONMOS1, LYONMOS2, SHIRAISHIHANLE, SASAKI}

In prior work, we have experimentally shown a substantial spin polarization suppression controlled by the attractive voltage bias on an electrostatically gated Si/SiO$_2$ interface in long-distance lateral-geometry devices making use of ballistic hot electron transport through metallic thin film contacts for spin injection and detection.\cite{LATERALPRL} Simultaneously, we observed a reduction in the mean and standard deviation of the spin transport time obtained through coherent spin precession measurements.\cite{LARMORPRBRC} This behavior is in contrast to vertical spin transport devices under similar conditions where a shorter transit time always results in a larger spin polarization because of the finite bulk lifetime.\cite{BIQINPRL} 

These counterintuitive results were analytically reproduced with a one-dimensional model in Ref. \onlinecite{LATERALPRL} by simulating the gate voltage dependence of both spin polarization and spin transport time measurements with a phenomenological effective spin lifetime which decreased with increasing gate voltage. Although the experimental data could be matched successfully with the results of this simple scheme, the detailed role of the Si/SiO$_2$ interface (and exposed air/Si top surface) in spin relaxation and its intrinsic spin lifetime distinct from the bulk could not be discerned. To accomplish this, it is necessary to use a two dimensional model\cite{DERYLATERAL} since the gate is expected to change the transverse proximity of transported spin-polarized electrons to the Si/SiO$_2$ interface, a detail which cannot be captured by any one-dimensional model.

In the present work, we develop a two-dimensional finite-differences model to simulate transport in electrostatically gated lateral spin transport devices. Hall measurements on ballistically injected electrons are used to provide input on the effects of interfacial momentum scattering affecting charge transport. We show that the strong spin polarization suppression accompanying a reduced mean transit time and uncertainty observed in spin transport experiments is indeed accounted for by a Si/SiO$_2$ interfacial spin lifetime of $\approx$ 1ns, smaller than the bulk Si value by at least two orders of magnitude.\cite{BIQINPRL} Furthermore, inclusion of spin relaxation at the exposed air/Si surface is shown to produce incompatible trends in the spin transport characteristics and is therefore inadequate to account for the dominant mechanisms responsible for our empirical observations.

Our model geometry is shown in Fig. 1(a). The lateral transport channel is 10$\mu$m thick, and the spin injector and detector contacts are 10$\mu$m long. The total lateral extent of our computational region is 100 $\mu$m. The equipotential gate voltage on the boundary of a 1 $\mu$m-thick SiO$_2$ dielectric $V_G$ and the accelerating voltage applied to the detector $V_D$ are first used as partial Dirichlet boundary conditions in the solution of Poisson's equation 

\begin{equation}
\nabla \epsilon \nabla \phi = \rho,
\label{POISSON}
\end{equation}

\noindent where the space charge density $\rho$ is zero in the dielectric but equal to the donor impurity density from full depletion (calculated from room-temperature resistivity of 4 k$\Omega$cm in the Si to match our device characteristics). The dielectric constant $\epsilon$ is also spatially dependent (=3.5 in the SiO$_2$ and =12 in the Si transport channel). The finite differences method solves for the spatially discretized electrostatic potential $\phi$ by a simple inversion of the matrix representation of the linear operator in Eqn. \ref{POISSON}, where Neumann boundary conditions (zero field normal to the boundary) are used on the boundaries not adjacent to the injector, detector, or gate equipotentials. An example electrostatic potential map calculated in this way for $V_D=$10V and $V_G=$2V is shown in Fig. \ref{FIG1}(b), where we have additionally included a conduction band discontinuity of 3.1 eV between Si and SiO$_2$.\cite{SIO2BARRIER}

Transport of a single electron spin component $s$ in this potential is then governed by the classical two-dimensional drift-diffusion-relaxation equation (valid for the relatively high temperatures to be simulated and in an effectively bulk regime where the mean-free-path is much shorter than the device lengthscale)

\begin{equation}
\frac{ds}{dt}=\nabla D \nabla s - \nabla [v s]-s/\tau+G,
\label{DRIFTDIFFEQ}
\end{equation}

\noindent where the spin generation source $G$ is nonzero only at the injector.  We can ignore the other two vector components of spin because the empirical spin precession results in a perpendicular magnetic field can be reduced to a transit-time distribution for just one component via Fourier transform.\cite{LARMORPRBRC, LATERALPRL} We therefore only need to determine the drift velocity vector field $v$ and the diffusion coefficient scalar field $D$. Because the bulk spin lifetime in Si is several hundred ns at the temperature range of interest\cite{BIQINPRL} and the effects we wish to model indicate much stronger spin relaxation than what is already known about the bulk, we set the spin lifetime $\tau$ to be finite only at either the top and bottom interfaces as explained below.

From the potential $\phi$, we calculate the electric field $\vec{\mathcal{E}}=-\nabla \phi$. This is then used to extrapolate a spatially-dependent drift velocity vector $\vec{v}$ from empirical electric-field- and temperature-dependent time-of-flight measurements\cite{CANALI}. The (charge) diffusion coefficient can then be calculated using Einstein's relation $D=\mu k_B T/q$ where $\mu=|v|/|\mathcal{E}|$ is charge mobility, $k_B$ is the Boltzmann constant, $q$ is the fundamental electron charge and temperature $T$=60K is chosen to match Ref. \onlinecite{LATERALPRL}.

\begin{figure}
\includegraphics[width=6.25cm, height=7.25cm]{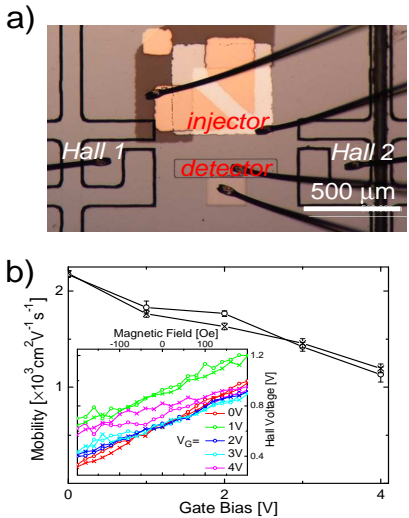}
\caption{\label{FIG2} 
(a) Plan view of 10-$\mu$m-thick lateral undoped Si spin transport device with Hall probes; (b) Example charge mobility directly measured with transport of ballistically injected electrons in a perpendicular magnetic field with $V_D=16V$ at 100K using Eqn. \ref{HALLEQN} and the data shown in the inset. The two sets of data shown here demonstrate the reproducibility and lack of any spurious hysteresis. This strong dependence on gate voltage motivates the interfacial mobility reduction incorporated into our finite differences model.  
}
\end{figure}

Whereas we can be confident in our application of bulk data to our model of the interior points of our Si channel, transport at the interfaces is however greatly modified by enhanced momentum scattering which reduces the mobility $\mu$.\cite{MOSMOBILITY} To determine the extent to which this occurs in our devices, we performed Hall measurements in magnetic field $B$ perpendicular to the device plane with contacts to the lateral transport region as shown in Fig. \ref{FIG2}(a). In contrast to ohmic transport, here with an undoped channel and rectifying Schottky contacts at the injector and detector, the carrier density is not constant but rather determined by the injected particle current density $J$ and the transport time $\bar{t}$ as $n=J\frac{\bar{t}}{L}$, where $L$ is the transport length. This gives a Hall voltage of 

\begin{equation}
V_{Hall}=Jw\frac{B}{n}=\frac{wLB}{\bar{t}}\approx B{\mu}\frac{w}{L}V_D, 
\label{HALLEQN}
\end{equation} 

\noindent where $w/L\approx 1$ gives the ratio of device width to transport length and we use the linear response approximation $L/\bar{t}\approx \mu V_D/L$. Hence, with ballistic hot electron injection into otherwise insulating lateral semiconductor channels as we have here due to low doping, low temperature, and rectifying Schottky barrier contacts,\cite{JANGMC} the Hall voltage is surprisingly independent of injection current to first order. Furthermore, in contrast to Hall measurements on ohmic systems which yield the carrier density, here the magnetic field dependence (slope vs $B$) gives the mobility $\mu$ directly. 

The example experimental mobility data from application of Eqn. \ref{HALLEQN} shown in Fig. \ref{FIG2}(b) give evidence for a strong momentum scattering enhancement upon application of gate voltage, with a reduction by a factor of 2 between $V_G=$0V and 4V. To account for this strong effect in our model, we need to lower the bulk drift velocity value in elements adjacent to the interfaces in the transport channel by a significant factor. However, because of the low doping density and relative lack of bandbending, our electrons are not confined entirely to the interface layer as in a two dimensional electron gas. Therefore, the mobility reduction measurement is an average over some electrons which never touch the interface and it is difficult to assign a precise value to the interfacial mobility reduction in the finite differences model based on experimental results. Measurements on similar (Si/SiO$_2$ field-effect) devices indicate a large characteristic mobility reduction of $\approx$10 in inversion layers due to e.g. interface roughness\cite{MOSMOBILITY, MOSMOBILITY2}; since this is consistent with our experimental observations, we therefore use that value in subsequent model calculations.

The drift-diffusion-relaxation equation in Eqn. \ref{DRIFTDIFFEQ} describes the evolution of spin density $s$ in time and space. To model the spin transit time in a straightforward way, we can simulate the sourceless evolution of an initial delta function spin distribution at the injector which can be done by a suitable technique explicit in time such as the Crank-Nicolson method.\cite{CHANPRBRC, SPINPLASMON} This corresponds to an iteration in time with stepsize $\Delta t$ (starting with a delta-function $\mathbf{s}^0$ nonzero only at the injector) of 

\begin{equation}
\mathbf{s}^{n+1}=(\mathbb{I}-\frac{\Delta t}{2}\mathbb{H})^{-1}\left[ (\mathbb{I}+\frac{\Delta t}{2}\mathbb{H}) \mathbf{s}^n\right],
\label{CN1DEQN}\end{equation}

\noindent where $\mathbf{s}^n$ is the array of spin density values at time $t=n\Delta t$. $\mathbb{H}$ is the block-tridiagonal \emph{matrix} representation of the drift, diffusion, and relaxation linear operators (with inhomogeneous $v$ and $D$) in the finite differences scheme and $\mathbb{I}$ is the identity matrix.  Elements at all interfaces are treated appropriate to the boundary conditions chosen; insulating boundaries are used (including at the Si/SiO$_2$ interface due to the conduction band offset barrier) except at the detector where $s=0$ so that spin flows out of the computational region to constitute a signal current.

Although explicit iteration in time as in Eqn. \ref{CN1DEQN} is a possible route to calculating the transit time, due to causality constraints $\Delta t$ must typically be chosen very small for numerical stability. Directly calculating relatively long timescale transport phenomena in this way with large matrices necessary to discretize two-dimensional space can therefore be computationally costly due to the large number of iterations required. 

\begin{figure}
\includegraphics[width=7.25cm, height=9.25cm]{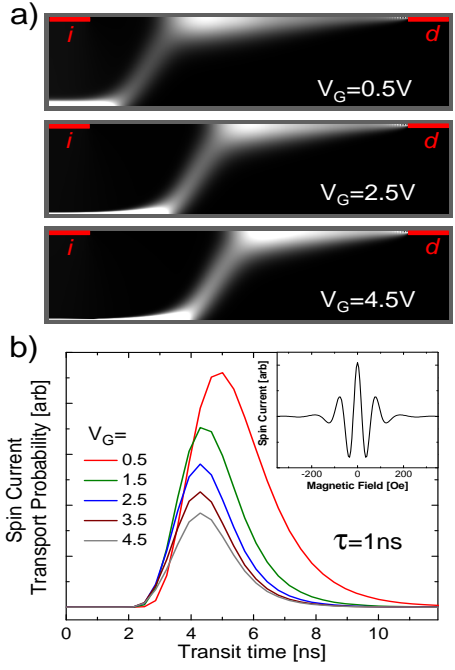}
\caption{\label{FIG3} 
(a) Example steady-state spatial spin distributions with constant spin source at injector ``i" obtained with matrix inversion of Eqn. \ref{MATRIXINVERSIONEQN} at $\omega=0$ for several gate voltages ($V_G$) with detector voltage $V_D=$10V; (b) Spin transit-time distributions upon arrival at detector ``d'' as a function of gate voltage at $V_D=$10V with Si/SiO$_2$ interfacial spin lifetime $\tau=$1 ns. Inset: Example spin precession simulation from which the spin current transport distributions are obtained via Fourier transform; here $V_G$=2V.
}
\end{figure}

Therefore, instead of using the Crank-Nicolson scheme in Eqn. \ref{CN1DEQN} (or other time-explicit alternatives), we transform Eqn. \ref{DRIFTDIFFEQ} into the frequency domain to obtain an equation for $\tilde{s}(\omega)=\int s(t) exp(-i\omega t) dt$. This allows us to directly simulate the steady-state response with spin precession in a perpendicular field (at frequency $\omega=g\mu_B B/\hbar$, where $g$ is the electron spin g-factor, $\mu_B$ is the Bohr magneton, and $\hbar$ is the reduced Planck constant). Within the finite differences scheme, we then must simply perform a matrix inversion to solve

\begin{equation}
(i\omega \mathbb{I}-\mathbb{H})\tilde{\mathbf{s}}=\mathbf{G}
\label{MATRIXINVERSIONEQN}
\end{equation}

\noindent where $\mathbf{G}$ is a constant source at the injector for each chosen value of $\omega$. Then, the time-domain response to an injected pulse can be reconstructed via the Fourier transform of the integrated transport current signal at the detector, $\tilde{J}_s(\omega)=-D\frac{d\tilde{s}}{dy}$.

Using discretization $\Delta x$=200nm and $\Delta y$=50nm, we have implemented the finite difference scheme outlined above on our spatial domain using sparse matrix operations in MATLAB. The results discussed below are insensitive to the choice of these discretization parameters so long as the meshing is detailed enough to prevent complications from spurious and unphysical oscillations which are well-known in finite-difference methods.\cite{FDOSCILLATION} In Fig. \ref{FIG3}(a), we show an example steady-state spatial distribution for $\omega=0$ at detector voltage bias $V_D$=10V and gate voltage biases $V_G$=0.5, 2.5, and 4.5V (both relative to the injector source potential). The injected spin is clearly attracted from the injector source at the upper left of each image to the electrostatically-gated Si/SiO$_2$ interface at the bottom of each image, and the electron distribution fills a larger lateral region along that interface as the gate voltage increases. Eventually, $V_D$ draws the electrons laterally to the detector, where it first passes through a region close to the exposed air/Si interface at the top of each image before being absorbed at the detector contact edge.

\begin{figure}
\includegraphics[width=7.25cm, height=12.25cm]{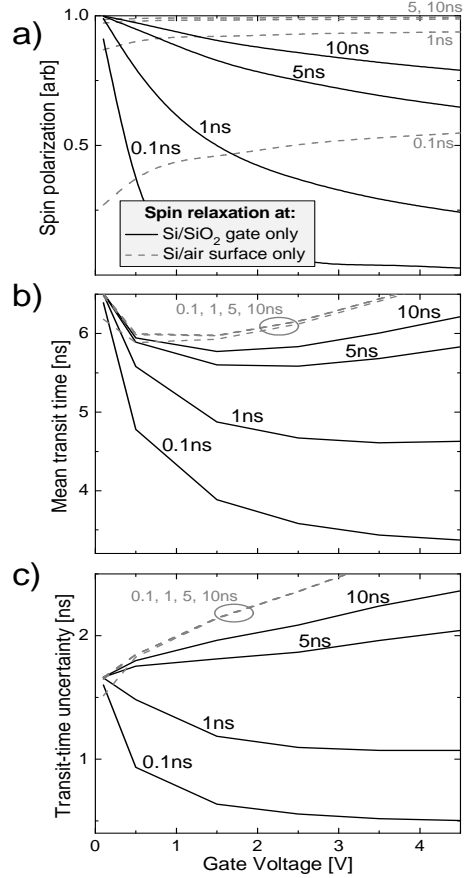}
\caption{\label{FIG4} 
Interfacial spin lifetime dependence of transport distribution properties as a function of gate voltage $V_G$. In (a), total spin polarization, (b) mean spin transit time, and (c) transit time uncertainty (distribution standard deviation). Black lines are simulations which include relaxation only at the Si/SiO$_2$ interface, whereas grey dashed lines include relaxation only at the air/Si top surface. Lifetimes of $\approx$1 ns are necessary to reproduce the experimentally-observed decreasing trends of all these parameters\cite{LATERALPRL}, in contrast to the known bulk lifetime in Si of several hundred ns at temperature of 60K.\cite{BIQINPRL}
}
\end{figure}

Fig. \ref{FIG3}(b) shows the spin current transit time distribution with a small interfacial lifetime $\tau=1$ns (only at the Si/SiO$_2$ backgate interface) reconstructed from simulated spin precession measurements by varying $\omega$ (as shown in the inset) for several gate voltages $0.5V<V_G<4.5V$. Clearly, as the injected electrons are attracted more to the backgate Si/SiO$_2$ interface with larger gate voltage, several changes occur: the total spin polarization (area under the curve normalized by the $\tau=\infty$ result to account for spin conservation) decreases, the mean transit time decreases, and the distribution width (transit time uncertainty leading to spin dephasing) decreases. All these trends mimic the empirical results from Ref. \onlinecite{LATERALPRL}.

This gate-voltage-dependent behavior is quantified by the distribution trends at different chosen Si/SiO$_2$ spin lifetimes $\tau=$0.1-10 ns shown by the solid lines in Fig. \ref{FIG4}(a)-(c). It can clearly be seen that for interfacial spin relaxation $\tau\apprge$ 1ns, the distribution parameters (spin polarization, mean transit time, and standard deviation) follow trends which are inconsistent with previous experimental observations in Ref. \onlinecite{LATERALPRL}: in particular, the latter two properties shown in Fig. \ref{FIG4}(b) and (c) \emph{increase} with increasing gate voltage when the lifetime is too long. 

If spin relaxation is instead included only at the top air/Si surface, the correspondence to experiment is even worse. As shown by dashed lines in Fig. \ref{FIG4}(a)-(c), the polarization, mean transit time, and transit time uncertainty are all predicted to \emph{rise} with applied gate voltage, regardless of the chosen interfacial spin lifetime, contradicting experiment. Therefore, our prior experimental results in Ref. \onlinecite{LATERALPRL} cannot be explained by top-surface spin relaxation at the air/Si interfoce, nor purely electrostatic effects. Adequate simulation requires the inclusion of dominant Si/SiO$_2$ interfacial spin relaxation at rates that are distinct in magnitude from the intrinsic mechanisms in the bulk which would otherwise lead to a spin lifetime orders of magnitude longer than the transit times seen here.\cite{YAFET}  

It should be noted that while this calculation is intended to capture the dominant transport mechanisms in thick-channel gated lateral devices, it is not an explicit simulation of the exact experiment in Ref. \onlinecite{LATERALPRL}. For instance, the injector-to-detector distance in the model is smaller to keep the calculation tractable and we do not attempt here to include the effect of spin injection only at the interior of the source contact as in true ballistic hot electron injection with a tunnel junction cathode. However, the features of the geometry are quite general, so modifications to account for these discrepancies are not expected to change the overall trends shown here. In addition, due to experimental unknowns such as interface dipole formation due to gate doping level and type, etc, we do not have the ability to precisely determine the exact empirical electrostatic boundary conditions from the experimental voltages applied, as evidenced by our ability to operate the device at negative gate voltage beyond where channel pinch-off conditions are expected.\cite{LATERALPRL} Our calculations here do however establish a consistent trend with increasing gate voltage which can only be explained by inclusion of a strong interfacial spin relaxation that becomes more and more dominant as the injected electrons are attracted to the Si/SiO$_2$ interface. 

Although we cannot yet directly identify the mechanism of spin relaxation at this Si/SiO$_2$ interface, the strong reduction in charge mobility as a function of gate bias shown in Fig. \ref{FIG2} suggests a contribution from the Elliott mechanism.\cite{ELLIOTT} In this case, because the two degenerate conduction band states are not pure spin states due to spin-orbit interaction,\cite{FABIANWU, DERYLI} momentum scattering -- even from spin-independent potentials -- can lead to spin flips. Closer proximity to the gate interface induced by rising gate voltage and its reduction in charge mobility from momentum scattering is therefore consistent with the lower interfacial spin lifetime we deduce from spin transport modeling. However, because the Elliott spin-flip rate is linearly related to the momentum scattering rate and hence inversely proportional to the mobility, a factor of two reduction in mobility over the gate bias range used here indicates this mechanism cannot be the dominant effect responsible for our observations of spin lifetime suppression of two orders of magnitude.  The role of other relaxation mechanisms such as exchange coupling to extrinsic paramagnetic spins (e.g. P$_b$ centers\cite{YABLONOVITCH}) and electric field fluctuations causing spin-orbit effects\cite{SHERMAN} must be explored. Although recently the effects of transient spin-polarized electron trapping in bulk defects have been observed,\cite{TRAPPING} the reduction in transit time with increasing bias seen in Ref. \onlinecite{LATERALPRL} is inconsistent with this mechanism playing a role here. It is hoped that further experimental efforts with e.g. spin resonance techniques on samples having deliberately prepared interfaces with variable defect density will be capable of clarifying the details of such relaxation pathways. 

This work was supported by the Office of Naval Research and the National Science Foundation. We acknowledge the support of the Maryland NanoCenter and its FabLab.


\begin{thebibliography}{28}
\expandafter\ifx\csname natexlab\endcsname\relax\def\natexlab#1{#1}\fi
\expandafter\ifx\csname bibnamefont\endcsname\relax
  \def\bibnamefont#1{#1}\fi
\expandafter\ifx\csname bibfnamefont\endcsname\relax
  \def\bibfnamefont#1{#1}\fi
\expandafter\ifx\csname citenamefont\endcsname\relax
  \def\citenamefont#1{#1}\fi
\expandafter\ifx\csname url\endcsname\relax
  \def\url#1{\texttt{#1}}\fi
\expandafter\ifx\csname urlprefix\endcsname\relax\def\urlprefix{URL }\fi
\providecommand{\bibinfo}[2]{#2}
\providecommand{\eprint}[2][]{\url{#2}}

\bibitem[{\citenamefont{Yafet}(1963)}]{YAFET}
\bibinfo{author}{\bibfnamefont{Y.}~\bibnamefont{Yafet}}, in
  \emph{\bibinfo{booktitle}{Solid State Physics-advances in Research and
  Applications}}, edited by
  \bibinfo{editor}{\bibfnamefont{F.}~\bibnamefont{Seitz}} \bibnamefont{and}
  \bibinfo{editor}{\bibfnamefont{D.}~\bibnamefont{Turnbull}}
  (\bibinfo{publisher}{Academic Press}, \bibinfo{address}{New York},
  \bibinfo{year}{1963}), vol.~\bibinfo{volume}{14}.

\bibitem[{\citenamefont{Cheng et~al.}(2010)\citenamefont{Cheng, Wu, and
  Fabian}}]{FABIANWU}
\bibinfo{author}{\bibfnamefont{J.~L.} \bibnamefont{Cheng}},
  \bibinfo{author}{\bibfnamefont{M.~W.} \bibnamefont{Wu}}, \bibnamefont{and}
  \bibinfo{author}{\bibfnamefont{J.}~\bibnamefont{Fabian}},
  \bibinfo{journal}{Phys. Rev. Lett.} \textbf{\bibinfo{volume}{104}},
  \bibinfo{pages}{016601} (\bibinfo{year}{2010}).

\bibitem[{\citenamefont{Li and Dery}(2011)}]{DERYLI}
\bibinfo{author}{\bibfnamefont{P.}~\bibnamefont{Li}} \bibnamefont{and}
  \bibinfo{author}{\bibfnamefont{H.}~\bibnamefont{Dery}},
  \bibinfo{journal}{Phys. Rev. Lett.} \textbf{\bibinfo{volume}{107}},
  \bibinfo{pages}{107203} (\bibinfo{year}{2011}).

\bibitem[{\citenamefont{Lepine}(1970)}]{LEPINE}
\bibinfo{author}{\bibfnamefont{D.}~\bibnamefont{Lepine}},
  \bibinfo{journal}{Phys. Rev. B} \textbf{\bibinfo{volume}{2}},
  \bibinfo{pages}{2429} (\bibinfo{year}{1970}).

\bibitem[{\citenamefont{Huang et~al.}(2007)\citenamefont{Huang, Monsma, and
  Appelbaum}}]{BIQINPRL}
\bibinfo{author}{\bibfnamefont{B.}~\bibnamefont{Huang}},
  \bibinfo{author}{\bibfnamefont{D.~J.} \bibnamefont{Monsma}},
  \bibnamefont{and}
  \bibinfo{author}{\bibfnamefont{I.}~\bibnamefont{Appelbaum}},
  \bibinfo{journal}{Phys. Rev. Lett.} \textbf{\bibinfo{volume}{99}},
  \bibinfo{pages}{177209} (\bibinfo{year}{2007}).

\bibitem[{\citenamefont{Huang and Appelbaum}(2010)}]{LARMORPRBRC}
\bibinfo{author}{\bibfnamefont{B.}~\bibnamefont{Huang}} \bibnamefont{and}
  \bibinfo{author}{\bibfnamefont{I.}~\bibnamefont{Appelbaum}},
  \bibinfo{journal}{Phys. Rev. B} \textbf{\bibinfo{volume}{82}},
  \bibinfo{pages}{241202} (\bibinfo{year}{2010}).

\bibitem[{\citenamefont{Demkov et~al.}(2005)\citenamefont{Demkov, Fonseca,
  Verret, Tomfohr, and Sankey}}]{SIO2BARRIER}
\bibinfo{author}{\bibfnamefont{A.~A.} \bibnamefont{Demkov}},
  \bibinfo{author}{\bibfnamefont{L.~R.~C.} \bibnamefont{Fonseca}},
  \bibinfo{author}{\bibfnamefont{E.}~\bibnamefont{Verret}},
  \bibinfo{author}{\bibfnamefont{J.}~\bibnamefont{Tomfohr}}, \bibnamefont{and}
  \bibinfo{author}{\bibfnamefont{O.~F.} \bibnamefont{Sankey}},
  \bibinfo{journal}{Phys. Rev. B} \textbf{\bibinfo{volume}{71}},
  \bibinfo{pages}{195306} (\bibinfo{year}{2005}).

\bibitem[{\citenamefont{Jang and Appelbaum}(2009)}]{LATERALPRL}
\bibinfo{author}{\bibfnamefont{H.-J.} \bibnamefont{Jang}} \bibnamefont{and}
  \bibinfo{author}{\bibfnamefont{I.}~\bibnamefont{Appelbaum}},
  \bibinfo{journal}{Phys. Rev. Lett.} \textbf{\bibinfo{volume}{103}},
  \bibinfo{pages}{117202} (\bibinfo{year}{2009}).

\bibitem[{\citenamefont{Behin-Aein et~al.}(2010)\citenamefont{Behin-Aein,
  Datta, Salahuddin, and Datta}}]{DATTANATURENANO}
\bibinfo{author}{\bibfnamefont{B.}~\bibnamefont{Behin-Aein}},
  \bibinfo{author}{\bibfnamefont{D.}~\bibnamefont{Datta}},
  \bibinfo{author}{\bibfnamefont{S.}~\bibnamefont{Salahuddin}},
  \bibnamefont{and} \bibinfo{author}{\bibfnamefont{S.}~\bibnamefont{Datta}},
  \bibinfo{journal}{Nature Nanotech.} \textbf{\bibinfo{volume}{5}},
  \bibinfo{pages}{266 } (\bibinfo{year}{2010}).

\bibitem[{\citenamefont{Dery et~al.}(2007)\citenamefont{Dery, Dalal, Cywinski,
  and Sham}}]{DERYNATURE}
\bibinfo{author}{\bibfnamefont{H.}~\bibnamefont{Dery}},
  \bibinfo{author}{\bibfnamefont{P.}~\bibnamefont{Dalal}},
  \bibinfo{author}{\bibfnamefont{L.}~\bibnamefont{Cywinski}}, \bibnamefont{and}
  \bibinfo{author}{\bibfnamefont{L.~J.} \bibnamefont{Sham}},
  \bibinfo{journal}{Nature} \textbf{\bibinfo{volume}{447}},
  \bibinfo{pages}{573} (\bibinfo{year}{2007}).

\bibitem[{\citenamefont{Dery et~al.}(2011)\citenamefont{Dery, Song, Li, and
  \v{Z}uti\'c}}]{SISPINCOMM}
\bibinfo{author}{\bibfnamefont{H.}~\bibnamefont{Dery}},
  \bibinfo{author}{\bibfnamefont{Y.}~\bibnamefont{Song}},
  \bibinfo{author}{\bibfnamefont{P.}~\bibnamefont{Li}}, \bibnamefont{and}
  \bibinfo{author}{\bibfnamefont{I.}~\bibnamefont{\v{Z}uti\'c}},
  \bibinfo{journal}{Appl. Phys. Lett.} \textbf{\bibinfo{volume}{99}},
  \bibinfo{pages}{082502} (\bibinfo{year}{2011}).

\bibitem[{\citenamefont{Xiao et~al.}(2004)\citenamefont{Xiao, Martin,
  Yablonovitch, and Jiang}}]{YABLONOVITCH}
\bibinfo{author}{\bibfnamefont{M.}~\bibnamefont{Xiao}},
  \bibinfo{author}{\bibfnamefont{I.}~\bibnamefont{Martin}},
  \bibinfo{author}{\bibfnamefont{E.}~\bibnamefont{Yablonovitch}},
  \bibnamefont{and} \bibinfo{author}{\bibfnamefont{H.}~\bibnamefont{Jiang}},
  \bibinfo{journal}{Nature} \textbf{\bibinfo{volume}{430}},
  \bibinfo{pages}{435} (\bibinfo{year}{2004}).

\bibitem[{\citenamefont{D'yakonov and Perel'}(1971)}]{DYAKONOV}
\bibinfo{author}{\bibfnamefont{M.}~\bibnamefont{D'yakonov}} \bibnamefont{and}
  \bibinfo{author}{\bibfnamefont{V.}~\bibnamefont{Perel'}},
  \bibinfo{journal}{Sov. Phys. Solid State} \textbf{\bibinfo{volume}{13}},
  \bibinfo{pages}{3023} (\bibinfo{year}{1971}).

\bibitem[{\citenamefont{Sherman}(2003)}]{SHERMAN}
\bibinfo{author}{\bibfnamefont{E.~Y.} \bibnamefont{Sherman}},
  \bibinfo{journal}{Appl. Phys. Lett.} \textbf{\bibinfo{volume}{82}},
  \bibinfo{pages}{209} (\bibinfo{year}{2003}).

\bibitem[{\citenamefont{Shankar et~al.}(2008)\citenamefont{Shankar, Tyryshkin,
  Avasthi, and Lyon}}]{LYONMOS1}
\bibinfo{author}{\bibfnamefont{S.}~\bibnamefont{Shankar}},
  \bibinfo{author}{\bibfnamefont{A.}~\bibnamefont{Tyryshkin}},
  \bibinfo{author}{\bibfnamefont{S.}~\bibnamefont{Avasthi}}, \bibnamefont{and}
  \bibinfo{author}{\bibfnamefont{S.}~\bibnamefont{Lyon}},
  \bibinfo{journal}{Physica E} \textbf{\bibinfo{volume}{40}},
  \bibinfo{pages}{1659 } (\bibinfo{year}{2008}).

\bibitem[{\citenamefont{Shankar et~al.}(2010)\citenamefont{Shankar, Tyryshkin,
  He, and Lyon}}]{LYONMOS2}
\bibinfo{author}{\bibfnamefont{S.}~\bibnamefont{Shankar}},
  \bibinfo{author}{\bibfnamefont{A.~M.} \bibnamefont{Tyryshkin}},
  \bibinfo{author}{\bibfnamefont{J.}~\bibnamefont{He}}, \bibnamefont{and}
  \bibinfo{author}{\bibfnamefont{S.~A.} \bibnamefont{Lyon}},
  \bibinfo{journal}{Phys. Rev. B} \textbf{\bibinfo{volume}{82}},
  \bibinfo{pages}{195323} (\bibinfo{year}{2010}).

\bibitem[{\citenamefont{Sasaki et~al.}(2010)\citenamefont{Sasaki, Oikawa,
  Suzuki, Shiraishi, Suzuki, and Noguchi}}]{SHIRAISHIHANLE}
\bibinfo{author}{\bibfnamefont{T.}~\bibnamefont{Sasaki}},
  \bibinfo{author}{\bibfnamefont{T.}~\bibnamefont{Oikawa}},
  \bibinfo{author}{\bibfnamefont{T.}~\bibnamefont{Suzuki}},
  \bibinfo{author}{\bibfnamefont{M.}~\bibnamefont{Shiraishi}},
  \bibinfo{author}{\bibfnamefont{Y.}~\bibnamefont{Suzuki}}, \bibnamefont{and}
  \bibinfo{author}{\bibfnamefont{K.}~\bibnamefont{Noguchi}},
  \bibinfo{journal}{Appl. Phys. Lett.} \textbf{\bibinfo{volume}{96}},
  \bibinfo{pages}{122101} (\bibinfo{year}{2010}).

\bibitem[{\citenamefont{Sasaki et~al.}(2011)\citenamefont{Sasaki, Oikawa,
  Shiraishi, Suzuki, and Noguchi}}]{SASAKI}
\bibinfo{author}{\bibfnamefont{T.}~\bibnamefont{Sasaki}},
  \bibinfo{author}{\bibfnamefont{T.}~\bibnamefont{Oikawa}},
  \bibinfo{author}{\bibfnamefont{M.}~\bibnamefont{Shiraishi}},
  \bibinfo{author}{\bibfnamefont{Y.}~\bibnamefont{Suzuki}}, \bibnamefont{and}
  \bibinfo{author}{\bibfnamefont{K.}~\bibnamefont{Noguchi}},
  \bibinfo{journal}{Appl. Phys. Lett.} \textbf{\bibinfo{volume}{98}},
  \bibinfo{pages}{012508} (\bibinfo{year}{2011}).

\bibitem[{\citenamefont{Dery et~al.}(2006)\citenamefont{Dery,
  Cywi\ifmmode~\acute{n}\else \'{n}\fi{}ski, and Sham}}]{DERYLATERAL}
\bibinfo{author}{\bibfnamefont{H.}~\bibnamefont{Dery}},
  \bibinfo{author}{\bibfnamefont{L.}~\bibnamefont{Cywi\ifmmode~\acute{n}\else
  \'{n}\fi{}ski}}, \bibnamefont{and} \bibinfo{author}{\bibfnamefont{L.~J.}
  \bibnamefont{Sham}}, \bibinfo{journal}{Phys. Rev. B}
  \textbf{\bibinfo{volume}{73}}, \bibinfo{pages}{041306}
  (\bibinfo{year}{2006}).

\bibitem[{\citenamefont{Canali et~al.}(1975)\citenamefont{Canali, Jacoboni,
  Nava, Ottaviani, and Quaranta}}]{CANALI}
\bibinfo{author}{\bibfnamefont{C.}~\bibnamefont{Canali}},
  \bibinfo{author}{\bibfnamefont{C.}~\bibnamefont{Jacoboni}},
  \bibinfo{author}{\bibfnamefont{F.}~\bibnamefont{Nava}},
  \bibinfo{author}{\bibfnamefont{G.}~\bibnamefont{Ottaviani}},
  \bibnamefont{and} \bibinfo{author}{\bibfnamefont{A.}~\bibnamefont{Quaranta}},
  \bibinfo{journal}{Phys. Rev. B} \textbf{\bibinfo{volume}{12}},
  \bibinfo{pages}{2265} (\bibinfo{year}{1975}).

\bibitem[{\citenamefont{Ohmi et~al.}(1991)\citenamefont{Ohmi, Kotani, Teramoto,
  and Miyashita}}]{MOSMOBILITY}
\bibinfo{author}{\bibfnamefont{T.}~\bibnamefont{Ohmi}},
  \bibinfo{author}{\bibfnamefont{K.}~\bibnamefont{Kotani}},
  \bibinfo{author}{\bibfnamefont{A.}~\bibnamefont{Teramoto}}, \bibnamefont{and}
  \bibinfo{author}{\bibfnamefont{M.}~\bibnamefont{Miyashita}},
  \bibinfo{journal}{IEEE Electron Device Lett.} \textbf{\bibinfo{volume}{12}},
  \bibinfo{pages}{652 } (\bibinfo{year}{1991}).

\bibitem[{\citenamefont{Jang and Appelbaum}(2010)}]{JANGMC}
\bibinfo{author}{\bibfnamefont{H.-J.} \bibnamefont{Jang}} \bibnamefont{and}
  \bibinfo{author}{\bibfnamefont{I.}~\bibnamefont{Appelbaum}},
  \bibinfo{journal}{Appl. Phys. Lett.} \textbf{\bibinfo{volume}{97}},
  \bibinfo{pages}{182108} (\bibinfo{year}{2010}).

\bibitem[{\citenamefont{ichi Takagi et~al.}(1994)\citenamefont{ichi Takagi,
  Toriumi, Iwase, and Tango}}]{MOSMOBILITY2}
\bibinfo{author}{\bibfnamefont{S.}~\bibnamefont{ichi Takagi}},
  \bibinfo{author}{\bibfnamefont{A.}~\bibnamefont{Toriumi}},
  \bibinfo{author}{\bibfnamefont{M.}~\bibnamefont{Iwase}}, \bibnamefont{and}
  \bibinfo{author}{\bibfnamefont{H.}~\bibnamefont{Tango}},
  \bibinfo{journal}{IEEE Trans. Electron Devices}
  \textbf{\bibinfo{volume}{41}}, \bibinfo{pages}{2357} (\bibinfo{year}{1994}).

\bibitem[{\citenamefont{Chan et~al.}(2009)\citenamefont{Chan, Hu, Zhang, Kondo,
  Palmstr\o{}m, and Crowell}}]{CHANPRBRC}
\bibinfo{author}{\bibfnamefont{M.~K.} \bibnamefont{Chan}},
  \bibinfo{author}{\bibfnamefont{Q.~O.} \bibnamefont{Hu}},
  \bibinfo{author}{\bibfnamefont{J.}~\bibnamefont{Zhang}},
  \bibinfo{author}{\bibfnamefont{T.}~\bibnamefont{Kondo}},
  \bibinfo{author}{\bibfnamefont{C.~J.} \bibnamefont{Palmstr\o{}m}},
  \bibnamefont{and} \bibinfo{author}{\bibfnamefont{P.~A.}
  \bibnamefont{Crowell}}, \bibinfo{journal}{Phys. Rev. B}
  \textbf{\bibinfo{volume}{80}}, \bibinfo{pages}{161206}
  (\bibinfo{year}{2009}).

\bibitem[{\citenamefont{Appelbaum et~al.}(2011)\citenamefont{Appelbaum, Drew,
  and Fuhrer}}]{SPINPLASMON}
\bibinfo{author}{\bibfnamefont{I.}~\bibnamefont{Appelbaum}},
  \bibinfo{author}{\bibfnamefont{H.~D.} \bibnamefont{Drew}}, \bibnamefont{and}
  \bibinfo{author}{\bibfnamefont{M.~S.} \bibnamefont{Fuhrer}},
  \bibinfo{journal}{Appl. Phys. Lett.} \textbf{\bibinfo{volume}{98}},
  \bibinfo{pages}{023103} (\bibinfo{year}{2011}).

\bibitem[{\citenamefont{Li et~al.}(2009)\citenamefont{Li, Tang, Warnecke, and
  Zhang}}]{FDOSCILLATION}
\bibinfo{author}{\bibfnamefont{J.}~\bibnamefont{Li}},
  \bibinfo{author}{\bibfnamefont{H.}~\bibnamefont{Tang}},
  \bibinfo{author}{\bibfnamefont{G.}~\bibnamefont{Warnecke}}, \bibnamefont{and}
  \bibinfo{author}{\bibfnamefont{L.}~\bibnamefont{Zhang}},
  \bibinfo{journal}{Math. Computation} \textbf{\bibinfo{volume}{78}},
  \bibinfo{pages}{1997–2018} (\bibinfo{year}{2009}).

\bibitem[{\citenamefont{Elliott}(1954)}]{ELLIOTT}
\bibinfo{author}{\bibfnamefont{R.~J.} \bibnamefont{Elliott}},
  \bibinfo{journal}{Phys. Rev.} \textbf{\bibinfo{volume}{96}},
  \bibinfo{pages}{266} (\bibinfo{year}{1954}).

\bibitem[{\citenamefont{Lu et~al.}(2011)\citenamefont{Lu, Li, and
  Appelbaum}}]{TRAPPING}
\bibinfo{author}{\bibfnamefont{Y.}~\bibnamefont{Lu}},
  \bibinfo{author}{\bibfnamefont{J.}~\bibnamefont{Li}}, \bibnamefont{and}
  \bibinfo{author}{\bibfnamefont{I.}~\bibnamefont{Appelbaum}},
  \bibinfo{journal}{Phys. Rev. Lett.} \textbf{\bibinfo{volume}{106}},
  \bibinfo{pages}{217202} (\bibinfo{year}{2011}).

\end{thebibliography}
\end{document}